\documentclass[aps,preprint,floatfix]{revtex4}
\usepackage{amsmath,amssymb,eucal,epsfig,subfigure}
\usepackage{color}
\begin{document}

\title{ PDB-NMA of a Protein Homodimer Reproduces Distinct Experimental Motility Asymmetry}

\author{Monique M. Tirion}
\email{mmtirion@clarkson.edu}
\author{Daniel ben-Avraham}

\affiliation{Physics Department, Clarkson University, Potsdam, New York
13699-5820, USA}

\begin{abstract}
We have  extended our analytically derived PDB-NMA formulation, ATMAN \cite{tirion14}, to include protein dimers using mixed internal and
Cartesian coordinates.
A test case on a 1.3\AA~resolution model of a small homodimer,  ActVA-ORF6, consisting of two  112-residue subunits  identically folded in a compact 50\AA~sphere,
reproduces the distinct experimental Debye-Waller motility asymmetry for the two chains, demonstrating that structure sensitively selects  vibrational signatures.
The vibrational analysis of this PDB entry, together with biochemical and crystallographic data, demonstrates the cooperative nature of the dimeric interaction of the
two subunits and suggests a mechanical model for subunit interconversion during the catalytic cycle.

\end{abstract}


\maketitle

\section{Introduction }


An internal force analysis of protein data bank (PDB) entries via normal mode analysis (NMA) reveals the softest coordinates inherent to the system and provides the basis of a theoretical derivation of the crystallographic Debye-Waller, or B factors. We have extended our PDB-NMA, ATMAN \cite{tirion14}, to dimeric protein systems according to the formalism presented in Levitt, Sander and Stern (LSS) \cite{LSS}. 
While completely general and valid for any collection of oscillators, our original implementation of the LSS formalism to protein systems was restricted to single chain polypeptides using only 
``soft''  torsional angle coordinates: the main chain $\phi$ and $\psi$ dihedrals as well as the side chains $\chi$ dihedrals \cite{tirion93}. For the analysis of dimers, we use these torsional angle coordinates
 $\{\phi_i^A,\psi_i^A,\chi_i^A;\phi_j^B,\psi_j^B,\chi_j^B\}$ for chain A and chain B, along with six additional coordinates, $\{x,y,z,\alpha,\beta,\gamma\}$, 
to account for the overall motion of chain B relative to chain A: the first three denote rigid body translations of chain B relative to A along the $x$-, $y$-, and $z$-direction; 
$\alpha$ is a rigid body rotation of chain B around its own $x$-axis (passing through B's center of mass), and $\beta$ and $\gamma$ are analogous rotations about B's $y$- and $z$-axes, respectively.

The LSS formalism extends straightforwardly to any kind of generalized coordinates, including mixed.  
Briefly, given a set of coordinates $\{q_l\}$ and a potential energy field $V(\{q_l\})$, one produces the Hessian matrix {\bf F},
\[
F_{ln}=\frac{\partial^2}{\partial q_l\partial q_n}V\;,
\]
and the inertia matrix {\bf H}:
\[
H_{ln}=\sum_{k=1}^N m_k\frac{\partial{\bf r}_k}{\partial q_l}\cdot\frac{\partial{\bf r}_k}{\partial q_n}\;,
\]
where the sum runs over all the atoms in the system ($m_k$ is the mass of the $k$-th atom and ${\bf r}_k$ is its  location vector).
The normal modes are then obtained by co-diagonalizing the {\bf F} and {\bf H} matrices:
\[
{\bf FA}={\bf\Lambda HA}\;.
\]
${\bf \Lambda}$ is a diagonal matrix containing the mode frequencies, $\Lambda_{ll}=\omega^2_l$, and $A_{kl}$ contains the $l$-th eigenmode amplitudes.  
(The modes are {\bf H}-orthogonal: ${\bf A}^T{\bf HA} = {\bf I}$.)  The $\{q_l\}$ in this formalism represent any kind of coordinates, including the mixed types we use in our analysis.  
The practical implementation, however, is far from trivial.  For example, the $\partial{\bf r}/\partial q$ derivatives required for the {\bf H}-matrix must exclude overall rigid body 
translations and rotations of the whole system, and derivatives of a complex potential energy field (needed for the computation of {\bf F}) require some care as well, especially 
since the different types of coordinates, coupled with the separate domains of chain A and chain B, give rise to numerous blocks in the {\bf F}-matrix, each with  its own set of rules, etc.  
We have tested our codes for the production of {\bf F} by performing all derivatives numerically (involving updates of the system's configuration by small increments $\delta q_i$), 
as well as analytically, and confirming the agreement between these two very different procedures.
In addition, the analyses  pass several  ``consistency" requirements: the diagonalization yields non negative eigenvalues; the emergent eigenfrequencies distribution is 
typical to folded proteins; and the root mean square deviations per mode~$i$, RMS${}^i$, are smoothly decreasing with $i$, with the first three modes obtaining in excess of 50\% of the total RMS.

One important reason to work with the complexity of dihedral angles is that these coordinates allow one to scale the computations to much larger  systems without sacrificing proper stereochemical
topology \cite{hayward95}. Dihedral bonds typically constitute a seventh or an eighth of the full complement of available degrees of freedom.   
The excluded degrees of freedom, bond lengths and bond angles, 
are quite stiffer than the soft dihedral coordinates and thus do not contribute to the slower modes of motion, the main focus of normal mode analyses.  
All-atom PDB-NMA that use Cartesian coordinates without topological constraints do not maintain proper stereochemistry and yield atypical eigenvalue spectra \cite{na2015,na2014}.
Design of proper topological constraints for another common reduction scheme, the use of only C$\alpha$ coordinates, is likewise challenging \cite{hinsen1998,bahar2005,jeong2006}.

Crystallized proteins adopt conformations that are long-lived and stable, implying that these structures already reside at a minimum of a multidimensional energy surface.
PDB-NMA therefore assumes a balanced distribution of pairwise atomic interactions, with suitably assigned  spring constants \cite{tirion96}.
Early formulations assigned a universal spring constant to all interatomic interactions, while current 
PDB-NMAs like sbNMA or ATMAN derive the pairwise interatomic spring constant strengths from the atom types and distances of separation of every interacting atom pair according
to a parent potential like CHARMM (sbNMA) \cite{na2014} or L79/ENCAD (ATMAN) \cite{tirion14}.  This yields identical results to those derived from classical NMA on energy minimized structures, when
both analyses proceed from the same, energy minimized structures.  The PDB data bank includes many structures with nearly identical folds: differences in their
vibrational signatures will be lost if such structures must be energy minimized first, motivating the use of
these types of energy potentials \cite{tirion15,tirion2017}.



To test our PDB-NMA formulation for dimeric protein systems,  we analyzed the PDB entry of a small homodimer with a large hydrophobic interface:
a monooxygenase  produced by a soil-dwelling bacterium, Streptomyces coelicolor.  
The enzyme, ActVA-Orf6, 
catalyzes the oxidation of large, 3-ringed aromatic polyketides in the biosynthetic pathway leading to actinorhodin production, one of four
 antibiotics produced by this gram-positive actinobacterium.
Sciara et al published the 1.3\AA~ resolution structure of the native, unliganded enzyme at 100K as PDB entry 1LQ9  in 2003 \cite{sciara2003}.


\begin{figure}[h]
\includegraphics[width=0.8\textwidth]{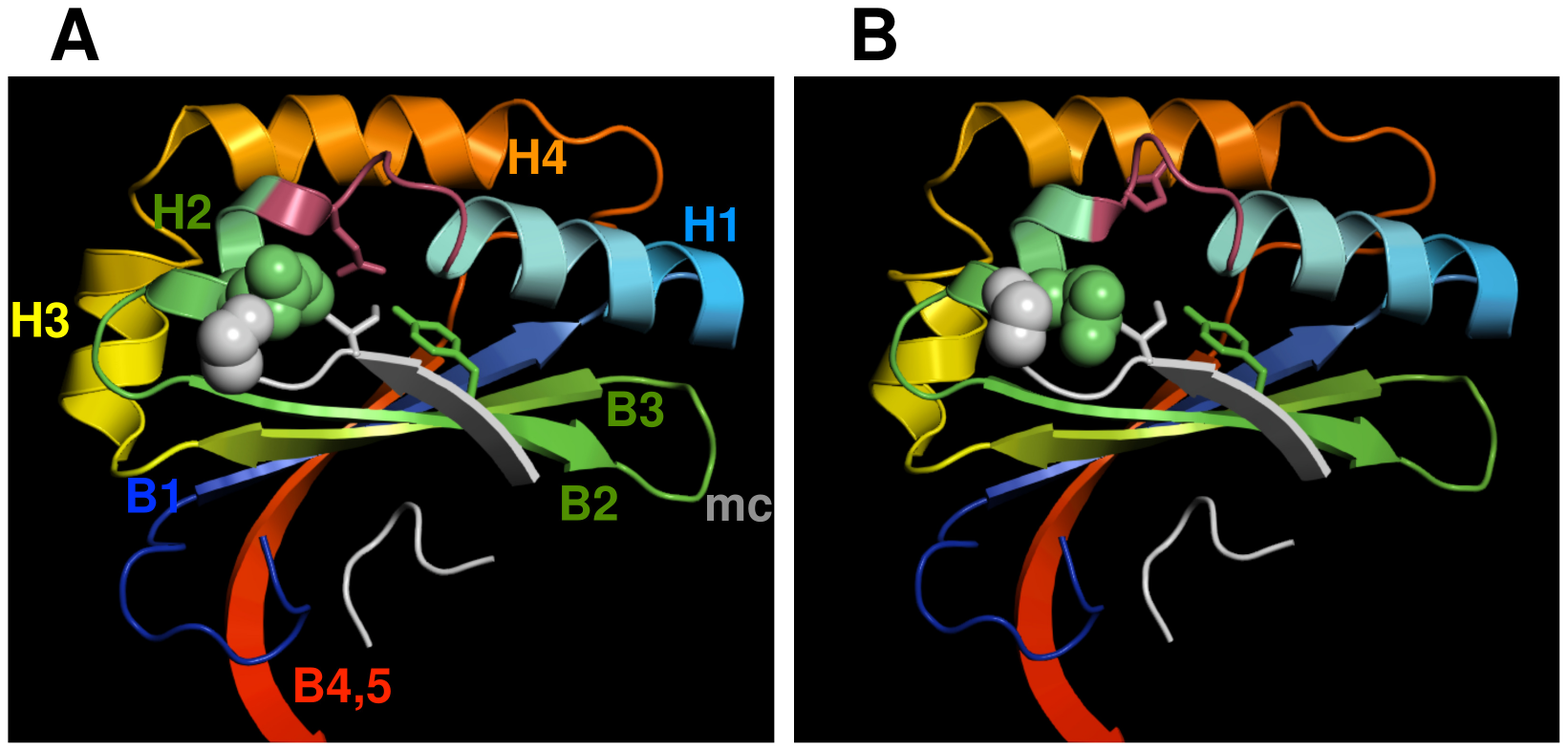}
\caption{Ribbon diagram of 1LQ9A (left) and 1LQ9B (RIGHT) using a rainbow coloring scheme from the N terminus (blue) to the C terminus (red), as
viewed from the `back' of the molecules.
The four $\alpha$ helices (H1-H4) and the four $\beta$ strands (B1-B4) are labeled. The final $\beta$ strand, B4, extends to form a fifth strand with the
adjoining chain's $\beta$ sheet at B2.  The adjoining chain's B5 from Pro106-Ser112, as well as the N terminal residues Ala1-Pro7 are added and colored in grey.
Tyr51 on B2, a likely catalytic residue, is drawn in green; the variably oriented Gln37 is drawn in burgundy along with the other residues in the 
asymmetrically mobile 34-38 loop; and the Ile110 peen of the adjoining subunit is drawn in grey. The mid chain region, Thr55-Th58, is labeled mc.
The distinct packing of the adjoining subunit's C terminal Ser113 (grey spheres) and Arg41 (green spheres) of the ERE pocket demonstrates how the
C terminal arm is latched when Gln37 is folded back (B)  or is unlatched when Gln37 is extended to form a hydrogen bond with the hydroxyl group of Tyr51.
The variable latching of the C terminal arms creates a variable mobility pattern that either  creates an enhanced swinging of the arched helices (B)
or a relative bouncing or hammering of Ile110 on Gln37 (A). As subunits A are seen to contain product-like analogues,  this mobility may 
enhance product release and possibly a rearrangement of Gln37 to the extended form (B) in preparation for entry of next substrate ligand.
}
\label{Figure01}
\end{figure}

 As seen in Fig. \ref{Figure01},
the enzyme consists of two identical chains of 
112 residues (Ala2-Ser113) with the secondary structure sequence
N-(B1-H1H2-B2)-(B3-H3H4-B4)-B5-C, that together fold into a compact sphere roughly 50\AA~in diameter.  
Each amino terminus N leads into  $\beta$ strand B1 and  loops back with a broken $\alpha$ helix H1H2 to form
the parallel $\beta$ strand B2.  Halfway through the primary sequence,  there  follows a sharp turn at Thr55-Thr58, and the same topology repeats:
$\beta$ strand B3, situated between and antiparallel to B1 and B2,  loops back via the broken $\alpha$ helix pair H3H4 to form a fourth antiparallel,
outer $\beta$ strand B4 aligned along B1.  Topologically, the $\beta$ strands present in the order B4, B1, B3 and B2, with the 
H1H2 helical arch over and parallel to B2 and next to it the H3H4 arch over and parallel to  B4.
This  fold presents as a $\beta$ sheet ``floor'' supporting two arches to create an open space accessible from the ``front'' (strand B4) as well as the ``rear'' (B2).
However, the C terminal sequence Phe103-Ile110 after B4 forms a final $\beta$ strand, B5, in the dimer. This $\beta$ strand extension of B4
  interlaces with and extends the neighboring chain's $\beta$ sheet at B2 and seals
 the rear aperture, creating an enclosed region, the active site cavity, accessible only from the front. The monomer's structure  is reminiscent of
a shell-shaped stage.

\begin{figure}[h]
\includegraphics[width=0.6\textwidth]{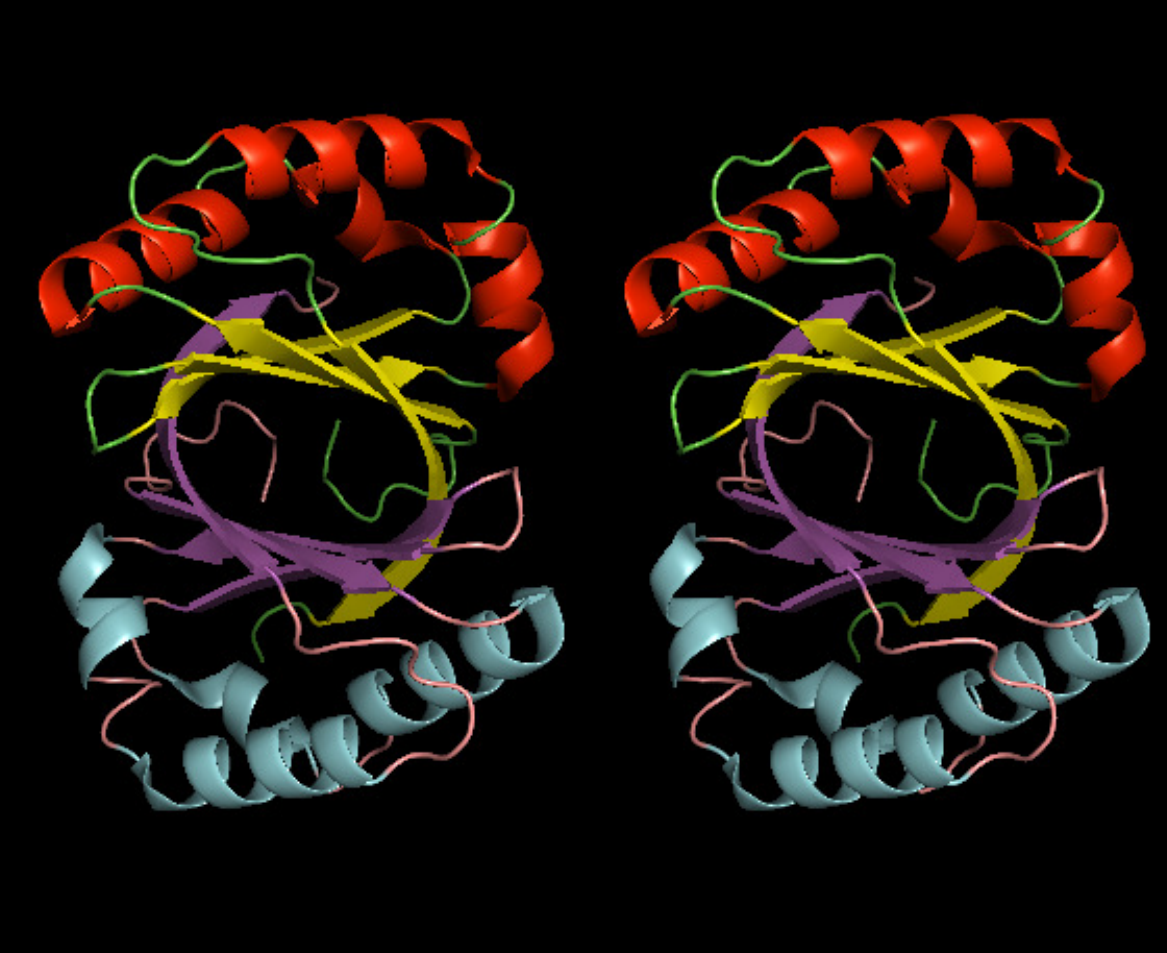}
\caption{Stereo ribbon-diagram of 1LQ9. Chain A at the top is in yellow/red/green while chain B is in magenta/cyan/pink.
In the dimer, the two $\beta$ sheets combine to form a barrel-shaped structure reminiscent of a TIM barrel. The N termini are visible at the rear of this
pseudo-barrel, forming a knob-like structure blocking solvent access to the barrel from the rear.
Two pairs of broken helices, H1H2 and H3H4 form arches over the $\beta$ sheet floors: the H3H4 arch framing the entrance to the active site in front and
the H1H2 arch behind. The rear of the active site is sealed by the neighboring subunit's C terminal arm.
The active site apertures are perpendicular to each other:  only one site seems to be active at a time.}
\label{Figure02}
\end{figure}

 In the dimerized complex, Fig. \ref{Figure02}, the two $\beta$ sheet floors are parallel, their main chains roughly 10\AA~apart and their  strands
juxtaposed by $90^\circ$, with the arched helices arrayed on the exterior surfaces, somewhat suggestive of a classic TIM barrel roll. 
Dimerization is mediated by the N- as well as the C-terminal regions of each chain. As mentioned, the two C terminal arms interlock with
the B2 strands of the opposing chain, zipper style, with seven hydrogen-bond ``teeth'' extending from the Phe103 carbonyl oxygen of one chain to the Thr55 amide hydrogen 
of the other chain, 
to the final interchain hydrogen bond between the Ile110 carbonyl oxygen of one chain to the Ala49 amide nitrogen of the other chain. 
As reported in Sciara \cite{sciara2003},
this swapped chain feature  results in a 2200\AA$^2$ contact area between the two chains. 
An additional  1100\AA$^2$ contact area is created by the interface of the N termini with  the Ala2Glu3Val4 juxtaposed in an antiparallel
fashion to create a
knob-like packing of the N termini  at the rear  of the central pseudo-barrel, between the C terminal arms. This packing arrangement  blocks access of solvent to the rear of the
barrel-type construct of the $\beta$ sheets. Altogether the 3300\AA$^2$  surface contact between the A and B chains represents  30\% of the total surface
area of the dimer, which is reported in the 1LQ9 header file as 11000\AA$^2$.  

A rigid-body, all-atom superposition of chains
A and B results in a RMSD of 0.5\AA, while a main chain superposition  results in a RMSD of 0.3\AA. The mainchain traces overlap closely, with 
the N termini as well as all five $\beta$ strands
 nearly perfectly aligned, while there exists a slight mismatch in the orientation of the loop (residues 34-38) linking H1 to H2 of the rear-most arch as well as a
slight overall shift of the H3 helix and break between helices H3 and H4 that frame the entrance to the active site cavity.
The C terminal arms after Pro98Pro99  again   align nearly perfectly to residue 109, 
with the final 4 C-terminal residues slightly out of register.  While most sidechains are seen to adopt identical conformations in chains A and B,
several sidechains situated on the surface and at the entry to the active site cavity orient  differently, including Gln37 which either orients towards the
active site cavity (chain A) or folds away from the cavity (chain B).

\section{Results}

Each chain consists of 845 heavy atoms and obtains 112 $\psi$ dihedrals, 101 $\phi$ dihedrals (the 10  proline $\phi$ dihedrals as well as the N terminal $\phi_1$ are fixed), and
165 $\chi$ sidechain dihedrals, or 378 dihedral coordinates per chain. Along with the 6 inter-monomer degrees of freedom therefore the dimer  obtains a total of 
762  dihedral coordinates versus 5064 Cartesian coordinates:
a nearly 7-fold reduction in coordinate space that leads to a significant reduction in matrix size. 

Chain A obtains 4609 intra-chain NBI and chain B 4666 intra-chain NBI between nonbonded atom pairs at least 4 bonds lengths apart 
and less than the cutoff distance defined by the inflection point of their
van der Waals curves~\cite{tirion14}; the atom pair forming the longest-range interaction having a distance of separation of 4.78\AA. 
The average number of NBI per atom for chain A is 9.74 and for chain B is 9.86.
The chains have zero cysteine residues and no disulfide bridges.  The number of inter-chain interactions between chains A and B, using identical selection criteria
as for the intra-chain NBI, is 714 of which 37 are 2-3\AA~ apart, 291 are 3-4\AA~ apart, and 386 are 4-5\AA~ apart.  

\begin{figure}[h]
\includegraphics[width=0.8\textwidth]{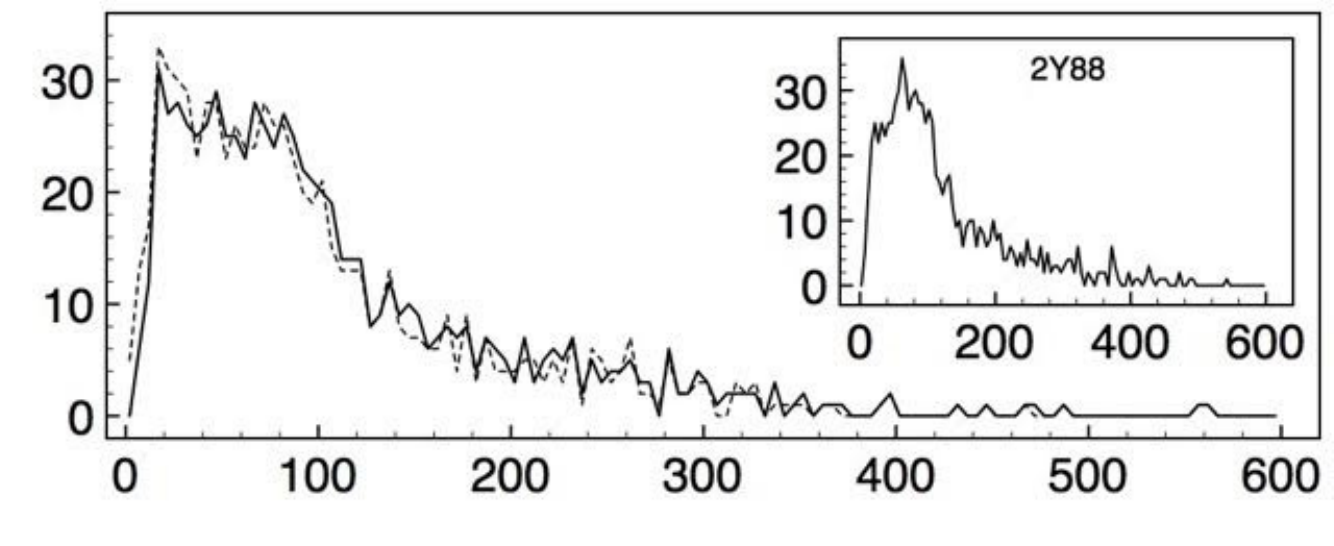}
\caption{The solid line presents the histogram of the eigenfrequencies of the 1LQ9 dimer with bin width of 5cm$^{-1}$, with the x-axis in units of cm$^{-1}$ and
the y-axis reporting the number of frequencies in this bin.
The dashed line
is the histogram of the eigenfrequencies of the isolated chain A summed with the histogram of the isolated chain B. 
The inset show the eigenfrequency
distribution of PDB entry 2Y88, which carries the signature of the universal character of singly folded protein chains.}
\label{Figure03}
\end{figure}

After matching the  analytically derived  to numerically computed terms of the dimer's Hessian matrix, we found that
the resultant eigenfrequencies vary smoothly from 5-600 cm$^{-1}$, with no anomalous or negative values.
The solid line in Fig. \ref{Figure03} shows the distribution of these frequencies as a histogram with bin width of 5cm$^{-1}$.
To ascertain the effects of dimerization on the eigenspectrum signature, we plot two additional curves.
The dashed curve is the histogram of the eigenfrequencies of the isolated chain A summed with the histogram of the isolated chain B (and therefore obtains
6 fewer modes). To compare these curves to the  `universal' curves obtained for singly folded protein chains~\cite{tirion93,Dani1,na2016universal}, we sought a PDB entry with similar
hallmarks as the dimer: a single chain consisting of  roughly 2$\times$112 residues folded in a TIM barrel fashion  
with a central eight-stranded $\beta$-barrel surrounded by a series of $\alpha$ helices.
We identified a high resolution 1.3\AA~ isomerase with
244 residues folded into  such a classic TIM barrel fold:  PDB entry 2Y88 \cite{due2011}.
The eigenspectrum distribution of this protein is shown in the inset of Fig. \ref{Figure03}. 

The inset plot for PDB entry 2Y88 demonstrates the typical or universal character of vibrational eigenspectra of single-chain crystalline proteins: 
a sparse concentration of slow modes 
rapidly rising to a densest concentration of modes in the 55-60cm$^{-1}$ region and then falling off with a broad shoulder  beyond 140cm$^{-1}$.
The sum of the isolated 1LQ9A and 1LQ9B spectra, shown as the dashed line in Fig. \ref{Figure03}, deviates from this pattern in having an anomalous concentration
of slow modes, pushing the maximum towards 20cm$^{-1}$. This is not surprising, as the isolated chains A and B represent unstable
configurations with unsupported  N- and C-terminal arms creating anomalous vibrational patterns.
Dimerization, it is seen in Fig. \ref{Figure03}, tightens these slowest modes with a decrease in the density in this region,
 but still retains a greater buildup of slow modes relative to isolated chains. This is not unexpected, as interfaces generally cannot
obtain a rigidity greater than their constituent bodies.

\begin{figure}[h]
\includegraphics[width=0.9\textwidth]{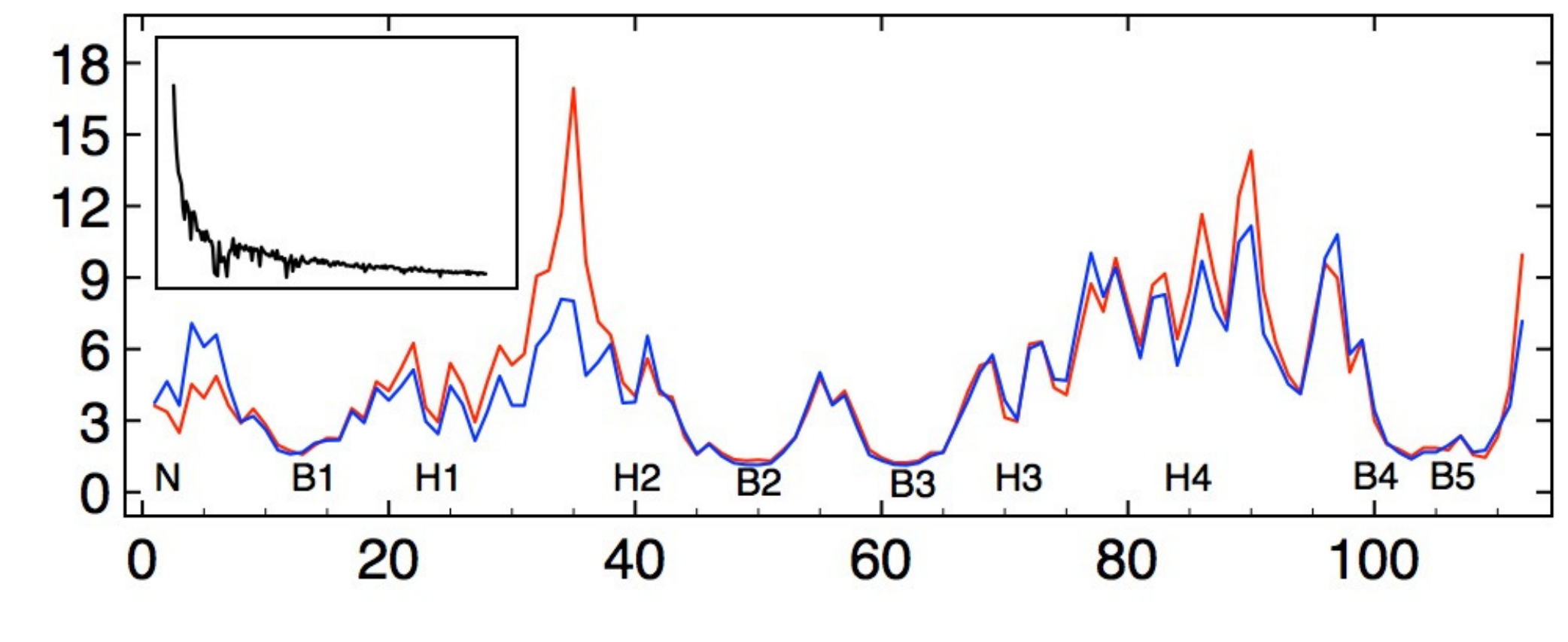}
\caption{The computed B factors of 1LQ9 in units of \AA$^2$ for chain A (blue) and chain B (red) for each of the C$_\alpha$ atoms.
The secondary structure elements are as indicated, with the H1H2 arch behind the H3H4 arch framing the  active site aperture.
The correlation of the curves is 0.91.
The inset shows the RMS$^i$ values for each of the first 200 eigenmodes; the y-axis extends from 0 to 0.15\AA.}
\label{Figure04}
\end{figure}

We next investigated the character of the eigenmodes  that give rise to these computed frequencies. The  RMS fluctuations of all $\alpha$ carbon atoms at room temperature
per mode $i$, RMS$^i$, are plotted in the inset of Fig. \ref{Figure04} for the first two hundred modes, with the y-axis extending from 0-0.15\AA.
The first three modes contribute 63\% to the  total 0.24\AA~(summed over all modes),
 while the higher frequency modes contribute roughly  0.02\AA~per mode, beyond the first 40 modes. This plot,  typical for protein systems,
gives a rough indication of the persistence length associated with each eigenmode: modes with longer correlation lengths
obtain larger RMS$^i$ values.

For PDB-NMA using simplified Hookean formulations, it is not unusual  for this plot to display an irregular decay in
amplitude, with one or more slow modes obtaining vanishingly small magnitudes.  
This irregularity is  absent when using standard classical potential energy formulations on energy-minimized structures.
An examination of such anomalous modes typically indicates an
inadequate ``knitting''  of the structure's motility, with some regions, such as surface side chains, moving independently of the rest of the structure, although
structural pathologies, such as steric clashes, will also give rise to this anomalous decay pattern.
ATMAN parameterizes every bonded and nonbonded interaction according to the parent potential, L79 \cite{L79}, and successfully reproduces results
from classical, energy-minimized NMA on single chains \cite{tirion14}. As the dimer interface of 1LQ9 obtains an extensive, hydrophobic and core-like packing
arrangement, no further adjustments to the nonbonded interaction list were required: 
the  RMS$^{i}$ plot  decays smoothly, as expected, for the slowest 40 modes, with some discontinuities appearing thereafter. 
As the current formulation ignores bond angle and bond length degrees of freedom
that likely contribute at these higher frequencies,  we report only on the slower modes.

The  computed B factors of each C$_\alpha$ due to the combined effect of the dimer's 40 slowest modes, unscaled and in units of \AA$^2$,
are shown in Fig. \ref{Figure04} for  chain A (blue) and  chain B (red). 
The five $\beta$ strands in each molecule obtain B values under 3\AA$^2$, reinforcing the description of these sheets as static floors
(a B factor of 5\AA$^2$ corresponds to a mean displacement from equilibrium of 0.25\AA).
The four $\alpha$ helices, the midpoint of the sequence at Thr55-Thr58,  as well as the C-terminal arm and the loop connecting the N terminal
``knob'' to B1  obtain B factor values over 5\AA$^2$,
with the break between H1 and H2 at the back of the stage and H4 plus the loop leading to B4 at the front especially mobile, with B values greater than 9\AA$^2$.

\begin{figure}[h]
\includegraphics[width=0.8\textwidth]{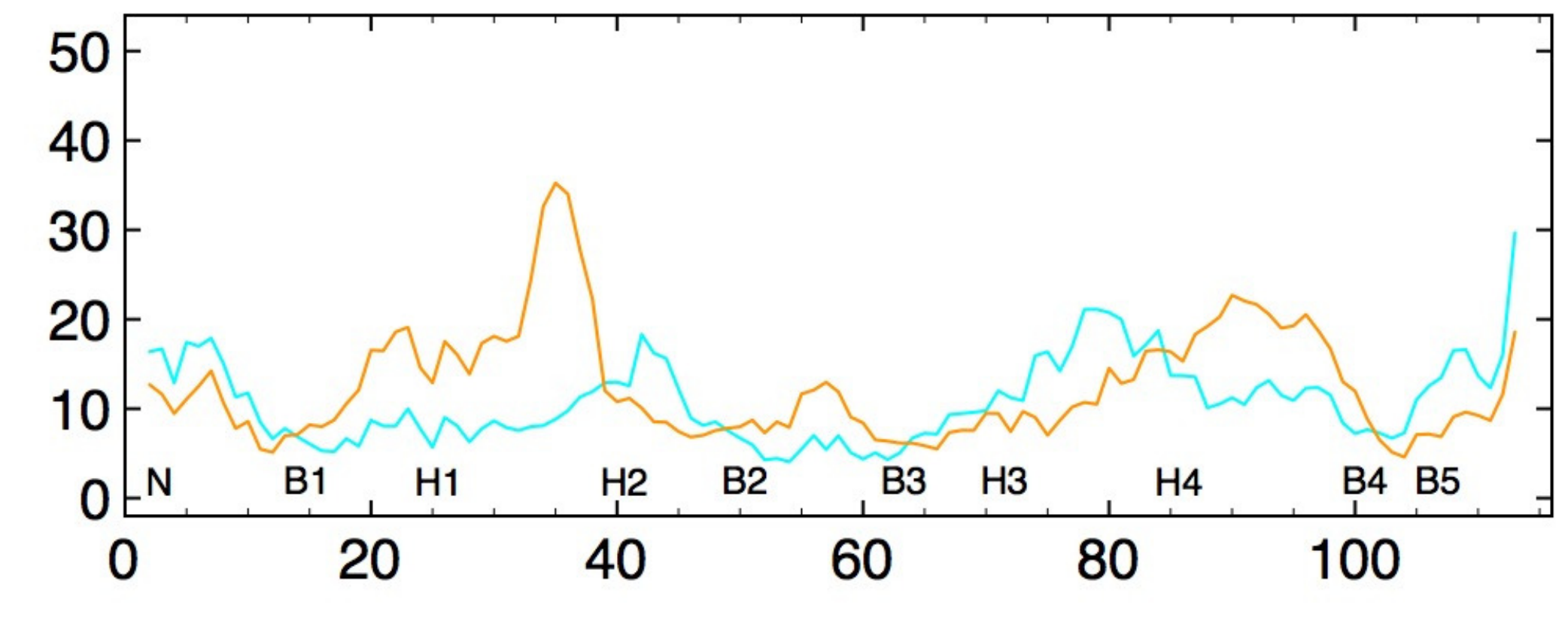}
\caption{The experimental B factors from PDB file 1LQ9 in units of \AA$^2$ for chain A (cyan) and chain B (orange) for each of the C$_\alpha$ atoms.
The correlation is 0.13.}
\label{Figure05}
\end{figure}

Fig. \ref{Figure04} demonstrates that
two identical subunits with a mainchain RMS overlap of 0.3\AA~do not necessarily obtain identical vibrational signatures. The computed B factors of each   C$_\alpha$ for chains A and B
obtain a correlation of 0.91, with residues 34-38 connecting H1 and H2 in subunit B  obtaining a peak B value
nearly twice that of chain A: 17\AA$^2$ versus 8\AA$^2$. In addition, the loop connecting N to B1  of chain A as well as the loop connecting H4 to B4 in
chain B appear to be more mobile than their partners.  Are these trends supported by the experimentally deduced B factors reported in the PDB file?
Fig. \ref{Figure05} shows the unscaled isotropic experimental B factors for each C$_\alpha$ atom for chain A (cyan) and chain B (orange). 
The experimental B values for chains A and B have a correlation of 0.13, with the motility of the arched helices H1H2 and H3H4 not well matched.

Experimental B factors obtain contributions not merely from intra-dimer, thermal vibrations, but also from numerous other sources including 
bulk solvent, inter-dimer (packing) vibrations, disorder such as mosaicity, and possible systematic errors \cite{Brunger1996}.
Generally, the non-vibrationary sources are homogeneous and provide a uniform noise level to the atomic coordinates.
Crystal vibrations span wavelengths delimited by the maximum crystal dimension  and the minimum unit cell dimension, and therefore  
contribute  a slowly varying ``noise'' within the unit cell. For this reason, variations in the B factors for the atoms comprising the asymmetric unit
are of  interest. Within the asymmetric unit, 
the relative contribution of plasticity (inter-minimum mobility) versus elasticity (intra-minimum mobility)
have been examined in an effort to quantify and characterize the structural heterogeneity within  protein crystals \cite{kuriyan1991,burling1994,levin2007,woldeyes2014,keedy2015}. 
 The current PDB-NMA demonstrates the degree to which a strictly elastic response of the reported structure to thermal perturbations
reproduces the experimentally deduced B factors.
And indeed, the experimental B factors in Fig. \ref{Figure05} broadly support the theoretical, intra-dimer predictions, with  
the $\beta$ strands obtaining smaller B values than either the helices, the midpoint Thr55-Thr58 or the N and
C terminal regions, and the distinct peak at 34-38 for chain B also evident.

\begin{figure}[h]
\includegraphics[width=0.9\textwidth]{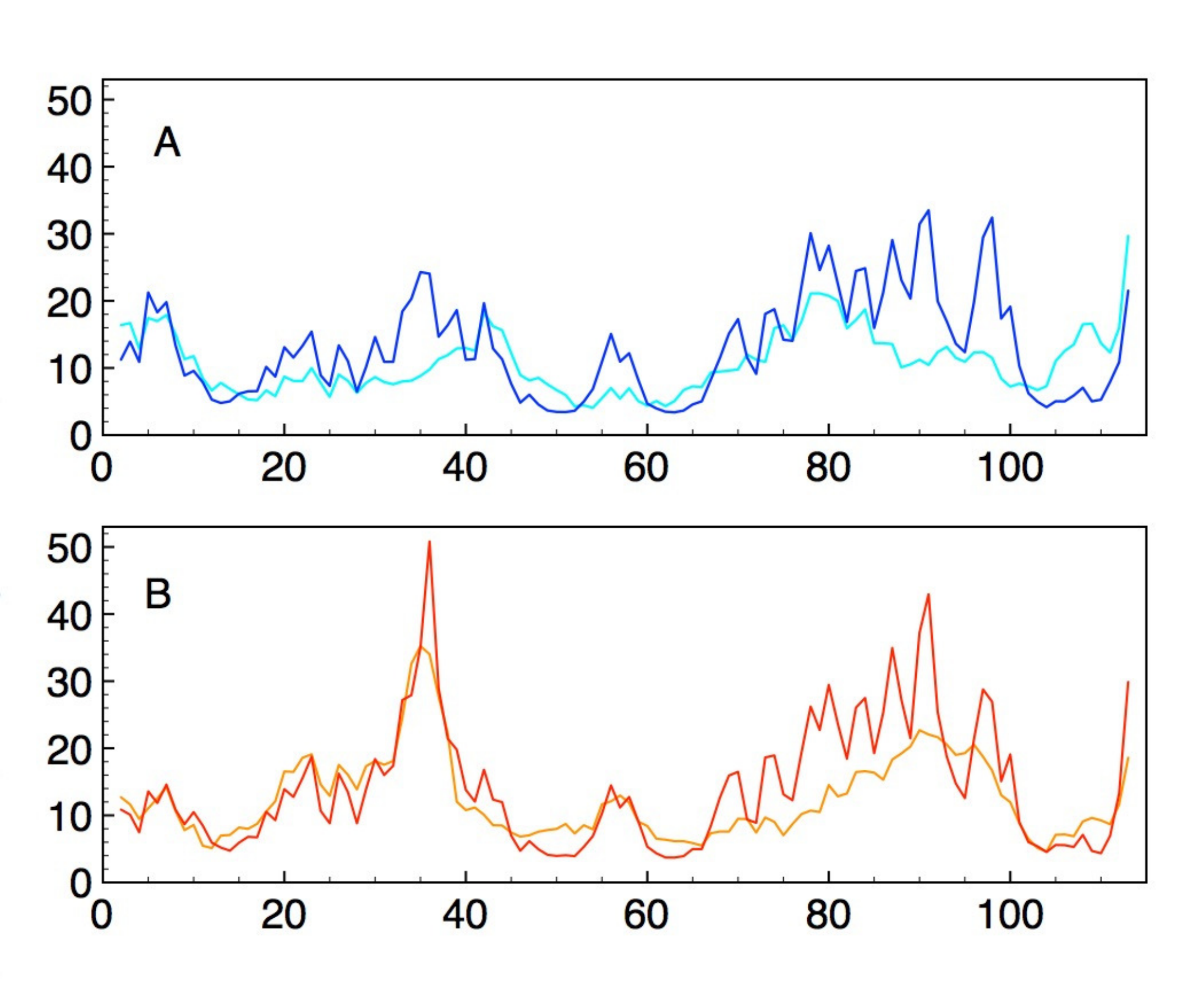}
\caption{A shows the experimental B factors (units of \AA$^2$) of the C$_\alpha$ atoms of chain A in cyan superposed with the scaled (by 3) theoretical B factors of the same chain in blue.
The correlation of the curves in A is 0.55. 
B shows the experimental B factors of chain B in orange superposed with the scaled theoretical B factors in red.
The correlation of the curves in B is 0.81.}
\label{Figure06}
\end{figure}

To more clearly ascertain the
overlap of the experimental and theoretical B factors,  we superpose the scaled theoretical B factors onto the experimental curves for chains A (Fig. \ref{Figure06}A)
and B (Fig. \ref{Figure06}B). A uniform scale factor of 3 applied to the computed values brings the minima of each curve into alignment in order to highlight the relevant correlations.
For this  PDB entry
the correlations  are reasonably high,
with the 40 slowest intra-dimer modes reproducing the observed
motility trends of the various secondary structure elements in each case.   

For chain A, the correlation of the experimental and computed B factors is 0.55,
with  over-estimates for the motilities of the  H1H2 arch, the midchain 55-58 loop, and the H4 to B4 region 
while  the C terminal region has a higher experimentally deduced motility.
Packing interactions may partly explain
these difference. 1LQ9 crystallized in an orthorhombic space group (P$2_12_12_1$) where each dimer obtains 4 unique packing interactions that introduce a total of
281 NBI (using identical selection criteria
as for the intra-chain NBI),  of which 6 are 2-3\AA~apart, 127 are 3-4\AA~apart, and 148 are 4-5\AA~apart.
These packing interactions, relatively few and weak in comparison to the intra- and inter-monomer interactions, are typically not included in PDB-NMA.
Visual inspection of the 1LQ9 model and its symmetrically related subunits shows that the region of closest approach 
   is between the external surface of the H3H4 arch of a B subunit that wedges into the  binding site aperture of an A chain. This packing
likely inhibits the vibrations in the 1LQ9 crystal of the front arch of chain A, and helps explain the lower observed motility of these regions.

Chain B obtains a  correlation of 0.81 between the experimental and computed B factors, 
including an excellent overlap for the 34-38 loop and close matches for the midchain loop and the N and C
termini. The greater motility predicted than observed for the H3H4 helical arch at the front of the ligand aperture is perhaps also due to the
packing arrangement that stacks the outer arches of dimers together.

\begin{figure}[h]
\includegraphics[width=0.9\textwidth]{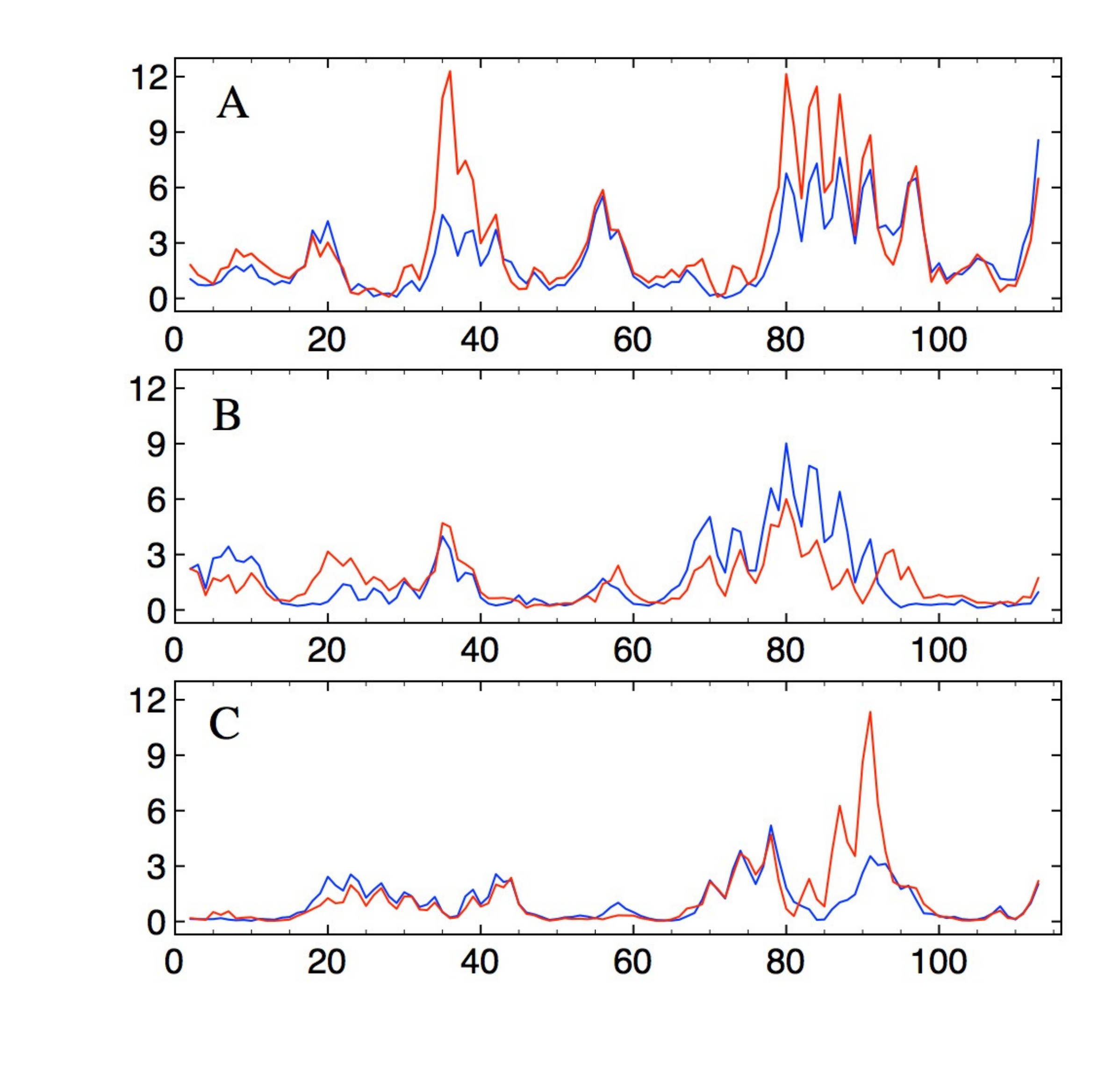}
\caption{The scaled (by 3) theoretical B factors in \AA$^2$ of the C$_\alpha$ atoms for subunits A (blue) and B (red) for modes 1 (A), 2 (B) and 3 (C).}
\label{Figure07}
\end{figure}

The relatively high correlations of the experimental to the theoretical B factors for chains A (0.55) and B (0.81), in contrast to the lower correlation of the experimental
curves comparing chains A and B (0.13), invites an examination of the character of those modes of the dimer that contribute to this asymmetry. As the slowest
3 modes contribute 63\% to the total computed RMS, we examined the character of each of those modes via 3d PyMol animations.
We plot the computed mean square motility, scaled to the experimental B factors, of each C$\alpha$ atom for modes 1, 2 and 3 and 
 for subunits A (blue) and B (red) in Fig. \ref{Figure07}A, B and C.

Mode 1 presents as a flexing or yawing of the two subunits, with the open, barrel-type construct between the two subunits (the front of the illustration in Fig \ref{Figure02})  
pulling apart 
and the C and N terminal regions at the opposing end of the barrel-construct  coordinating this flexing while maintaining, along with the Thr55-Thr58 midchain motif,
the  structural integrity of the dimer.  
The contribution of $\beta$ strands B5 to mode 1 is distinct, with the 7 hydrogen bonds linking each B5 strand with the opposing chain's B2 strand strained, but in a non-equivalent manner.
AB5 remains in register with BB2 while straining the interstrand bonds laterally away from the plane of the $\beta$ sheet floor, while the interstrand distance of separation
remains fixed between BB5 and AB2 as those two strands display a tendency to slide past each other. The source of this variable flexibility appears to be related to
 the manner in which each C terminal packs Ser113.  
In subunit B, Gln37 folds away from the binding cavity towards the subunit interface,  which simultaneously extends Arg41 to interact with ASer113, 
latching this C terminal residue, along with Glu38 and Glu41, in a ``ERE pocket'' (Fig. \ref{Figure01}).
 In subunit A, in contrast, Gln37 extends away from the interface toward the binding cavity, Arg41 is pulled away from  BSer113,  
the ERE pocket does not form and BSer113 is not similarly latched but is instead  solvent exposed.
 This variable packing of the C termini results in a variable transmission of motility, with the
AB5 strand, latched at ASer113 to the neigboring ERE pocket in BH2, vibrating in sync with that $\alpha$ helix and thence also the neighboring BH4 region.
Meanwhile, the unlatched and less restrained BB5 strand swings out of tandem with AH2, hammer-style, with its Ile110 sidechain like a peen striking  the extended AGln37 sidechain.
So for example, the distance of separation between CD1 of BIle110 and OE1 of AGln37 in the crystal structure is 2.8 \AA~
and varies from 2.6\AA~ to 3.1\AA~ due to mode 1, while simultaneously, at the other subunit, the distance of separation between the CD1 of AIle110 and OE1 of BGln37 remains at a steady 8.2\AA. 
This feature, then, explains the asymmetry between the computed  B factors of chains A and B of Fig. \ref{Figure07}A: when yoked to the neighboring subunit's H2 motif at the ERE pocket, 
the vibrational motility  of the C terminal $\beta$ strand adds to that of the H2 helix and thence also to H4, via the base stacking of 
Trp39 with Phe76 for example, and numerous other noncovalent interactions between the H1H2 and H3H4 arches.  The C terminal Ser113 that does not latch into the
ERE pocket  does not vibrate in tandem with H2 and does not contribute as effectively to that region's displacement in this mode of oscillation.


Higher order modes, it will be seen, reveal various motility patterns of the arched helices, but only mode 1 reveals a
high motility of the C terminal arms and the midchain peaks as well as the distinct chain A versus B asymmetry at H2 (Fig. \ref{Figure07}A). 
In higher order modes the B5 strands oscillate in tight synchrony with the
B2 strands of the opposing chain, no yawing of the barrel-construct between the chains is apparent, and the midchain peaks are likewise damped.
The source of the experimental H2 B factor  asymmetry, therefore, may be due to a single mode, the softest dimer mode,
and the variable packing of the C termini which either effectively transmits the intrinsic vibrational propensities to H2, or not.

Mode 2 presents as a sidewise tilting of the dimer that does not distort the central-barrel motif, but instead swings the active site roofs, especially
 the front arches H3H4, forward. This
motility seems to be mediated by the B4B5 junction with Pro106 serving as a hinge to allow the relative reorientation of B4 and B5.
While the character of the motility at the two active site cavities is identical,  their amplitudes of activation are not. This asymmetry is likely due to the variable packing
within the active site cavities  of residues such as Glu37 at the rear and Arg86 at the entrance of the active site.
 In subunit B, for example, Arg 86 is fully extended and directed into the active site cavity, while in subunit A it folds away from
the active site cavity, towards the exterior surface of the protein. 
The result is that the vibrational character of mode 2 is  differently expressed in subunits A and B, with the packing arrangement
within the active site of subunit B seeming to frustrate the motility expressed in subunit A.

Mode 3 presents as a type of twisting of the two chains resulting once again in a high mobility of the arches, H3H4, framing the active site cavity.
Rather than tilt forwards as in mode 2, in mode 3 the arches are seen to tilt sidewise in a manner that distorts the shape of the active site entrance.
Interestingly,  the vibrational character of mode 3 seems to favor subunit B, with BH4 and the subsequent loop to BB4 obtaining noticeably larger
amplitudes of vibration than in subunit A (Fig. \ref{Figure07}C).  It is interesting to note that
Pro80 and Pro93 bracket $\alpha$ helix H4 while Pro93 and Pro98Pro99 similarly bracket the loop connecting H4 to B4.
Prolines disrupt mainchain hydrogen bonding patterns and also reduce the instrinsic flexiblity of the polypeptide chain by eliminating the $\phi$ rotational degree
of freedom, and thus are well suited to isolate regions with flexibility characteristics distinct from neighboring regions, as seen in mode 3.

\section{Discussion}

The PDB-NMA of entry 1LQ9 contributes new insights into the structure/function relationship of ActVA-Orf6. 
The crystallographically determined atomic model implicates two gates for the reaction mechanism for this homodimeric enzyme \cite{sciara2003}.
A proton gate consisting of Arg86 at the entrance to the active site, together with
Tyr72 inside the active site cavity, are thought to shuttle the proton extracted from  substrate to bulk solvent.  
A  water gate consisting of Gln37 in the mobile 34-38 loop opens a narrow passage for diffusion of molecular oxygen from the bulk solvent or of a water molecule to the bulk. 
As evidence, Sciara et al point out that
only the extended Gln37 conformation in subunit A forms a hydrogen bond with Tyr51, a residue implicated in catalysis, whereas in subunit B  this residue
folds back and a water molecule replaces Gln37 in its interaction with the hydroxyl of Tyr51.
The crystal model \cite{sciara2003} together with biochemical data \cite{kendrew1997} also demonstrate that disruption of the dimeric interface eliminates catalytic activity: the enzyme
functions only as a dimer. In addition, the crystal structures of the enzyme with different ligands demonstrate that in the crystals, subunits A seem to bind product-like analogues
while subunits B seem to bind substrate-like analogues \cite{sciara2003}.

Crystallographically deduced B factors reveal that the two chains of this homodimer
obtain distinct mobility signatures for the H1H2 and H3H4 arches, with especially the mobile 34-38 loop connecting H1 and H2
 in subunit B obtaining significantly higher motility than in subunit A.
Without a vibrational analysis of the proposed structure, however, the source of the variation and asymmetry of the B factors cannot be discerned.
{We find that the 
PDB-NMA of the 1LQ9 model reproduces the  motility patterns for chains A and B,    suggesting that the  distinct B factor signatures are not an
artefact of crystal packing interactions but rather an innate feature of the chains.} An examination of the  modes of oscillation intrinsic to this homodimer 
provides a mechanistic explanation of the B factors that implicates a special role for the C terminal $\beta$ strands.

Taken together, the biochemical, crystallographic and vibrational analyses demonstrate the cooperative nature of the interaction of the two subunits in this homodimer.
Subunit A, bound with product and with Gln37 extended to interact with Tyr51 in the active site, retracts Arg41 and does
not form the ERE pocket to restrain the swinging of subunit B's C terminal arm. The resultant  oscillations of the C terminal hammer has  the Ile110 peen
 strike Gln37,  which serves perhaps to dislodge product and simultaneously to reorient the Gln37 residue to the retracted orientation compatible with 
substrate binding. The retraction of Gln37 simultaneously extends Arg41 that therefore snags the C terminal arm of the other subunit, whose vibrational motility  then contributes
to the  vibrational propensity of the arched helical roof. 
This sequence of events would sequentially interconvert subunit A to subunit B. The motilities presented in mode 1 appear consistent with 
the proposed Gln37 water gate, while modes 2 and 3 might serve to expedite entry and exit of ligands.
Crystal structures reveal only dimers with distinct subunits, A and B, and never  with  identical orientations of both subunits,  suggesting that
the chains catalyze substrate sequentially. How this higher level of cooperativity might be achieved is unclear, although it might involve the
tight dynamical coupling and rich communication between the C terminal arms across the N terminal knobs.
The importance to enzymatic activity of residues that confer specific dynamic signatures, such as Ser112, Arg41, Pro112 or Pro106, could be tested by
site directed mutagenesis studies.

\section{Conclusion}

The formalism  for macromolecular NMA  presented in LSS \cite{LSS} is completely general and applies to multimeric systems using mixed, internal and Cartesian, coordinates.
An application of PDB-NMA to a high resolution homodimer provides a plausible representation of internal flexibility of the complex, and requires no special reformulation
of the inter-protein potential energy expression presented in Tirion \& ben-Avraham \cite{tirion14}.
The reliability of the computation for this multimeric system is demonstrated by internal consistency checks that include matching the numeric and analytic
expressions of each term in the Hessian matrix; a diagonalization of the generalized eigenvalue equation that results in no negative eigenvalues;
an eigenvalue histogram distribution that adheres to the universal character of stably folded proteins; and an RMS$^i$ curve that smoothly decreases as
a function of mode number. In addition to these internal consistency checks, the computed B factors exhibit a high degree of correlation to the
experimentally deduced B factors, including the presence of a distinct peak in one of the chains. 
The precise pattern of covalent and noncovalent bonding deduced from crystallographic diffraction data therefore selects unique
vibrational signatures consistent with the independently refined Debye-Waller factors. 

 Vibrational analyses of PDB entries that
maintain stereochemically refined bond lengths and bond angles and adhere to the character of  bonded and nonbonded pairwise
atomic interactions as provided by standard energy potentials are feasible, both for monomeric and multimeric systems.
PDB-NMA is a quick, concise and reproduceable way to characterize the vibrational propensities of stably folded protein systems,
and provide a rich source of insight to further comprehend and model enzymatic activities.


\begin{thebibliography}{99}

\bibitem{tirion14}
 M.~M.~Tirion and D.~ben-Avraham, 
 ``Atomic torsional modal analysis for high-resolution proteins,'' 
 {\it Phys. Rev. E} {\bf91}, 032712 (2015).
 
\bibitem{LSS} 
 M.~Levitt, C.~Sander, and P.~S.~Stern, 
 ``The normal modes of a protein: Native bovine pancreatic trypsin inhibitor,"
 {\it Int. J. Quantum Chem.} {\bf24}, 181Ð99 (1983).
 
\bibitem{tirion93} 
 M.~M.~Tirion and D.~ben-Avraham, 
 ``Normal mode analysis of g-actin,"
 {\it J. Mol. Biol.} {\bf230}, 186195 (1993).
 
 \bibitem{hayward95} 
 Steven Hayward and Nobuhiro Go, 
 ``Collective variable description of native protein dynam- ics,''
 {\it Annual review of physical chemistry} {\bf46}, 223Ð250 (1995).
 
\bibitem{na2015} 
 Hyuntae Na and Guang Song, 
 ``Conventional nma as a better standard for evaluating elastic network models,''
 {\it Proteins: Structure, Function, and Bioinformatics} {\bf83}, 259Ð267 (2015).
 
\bibitem{na2014} 
 Hyuntae Na and Guang Song, 
 ``Bridging between normal mode analysis and elastic network models,''
 {\it Proteins: Structure, Function, and Bioinformatics} {\bf82}, 2157Ð2168 (2014).
 
\bibitem{hinsen1998} 
 Konrad Hinsen, 
 ``Analysis of domain motions by approximate normal mode calculations,''
{\it Proteins-Structure Function and Genetics} {\bf33}, 417Ð429 (1998).
 
\bibitem{bahar2005} 
 Ivet Bahar and AJ Rader, 
 ``Coarse-grained normal mode analysis in structural biology,''
 {\it Current opinion in structural biology} {\bf15}, 586Ð592 (2005).
 
\bibitem{jeong2006} 
 Jay I Jeong, Yunho Jang, and Moon K Kim, 
 ``A connection rule for carbon coarse-grained elastic network models using chemical bond information,''
 {\it Journal of Molecular Graphics and Modelling} {\bf24}, 296Ð306 (2006).
 
\bibitem{tirion96} 
 M.~M.~Tirion, 
 ``Large amplitude elastic motions in proteins from a single-parameter, atomic analysis,''
 {\it Phys. Rev. Lett.} {\bf77}, 1905 (1996).
 
\bibitem{tirion15} 
 M.~M.~Tirion,  
 ``On the sensitivity of protein data bank normal mode analysis: An application to GH10 xylanases,''
 {\it Phys. Biol.} {\bf12}, 066013 (2015).
 
\bibitem{tirion2017} 
Monique M Tirion, 
``A comparison of the innate flexibilities of six chains in f1-atpase with identical secondary and tertiary folds; 3 active enzymes and 3 structural proteins,''
{\it Structural Dynamics} {\bf4}, 044001 (2017).

\bibitem{sciara2003} 
Giuliano Sciara, Steven G Kendrew, Adriana E Miele, Neil G Marsh, Luca Federici, Francesco Malatesta, Giuliana Schimperna, Carmelinda Savino, and Beatrice Vallone, 
``The structure of actva-orf6, a novel type of monooxygenase involved in actinorhodin biosynthesis,''
{\it The EMBO journal} {\bf22}, 205Ð215 (2003).

\bibitem{Dani1} 
D.~ben-Avraham, 
``Vibrational normal-mode spectrum of globular proteins,''
{\it Phys. Rev. B} {\bf47}, 14559 (1993).

\bibitem{na2016universal} 
H.~Na, G.~Song, and D~ben-Avraham, 
``Universality of vibrational spectra of globular proteins,''
{\it Phys. Biol.} {\bf13}, 016008 (2016).

\bibitem{due2011} 
Anne V Due, Jochen Kuper, Arie Geerlof, Jens Peter von Kries, and Matthias Wilmanns, 
``Bisubstrate specificity in histidine/tryptophan biosynthesis isomerase from mycobacterium tuberculosis by active site metamorphosis,''
{\it Proceedings of the National Academy of Sciences} {\bf108}, 3554Ð3559 (2011).

\bibitem{L79} 
M.~Levitt, 
``Molecular dynamics of native protein. I. Computer simulation of trajectories,''
{\it J. Mol. Biol.} {\bf168}, 595Ð620 (1983).

\bibitem{Brunger1996} 
Axel T.~Br{\"u}nger, 
``Recent developments for crystallographic refinement of macromolecules,''
in {\it Crystallographic Methods and Protocols}, edited by Christopher Jones, Barbara Mulloy, and Mark R. Sanderson (Humana Press, Totowa, NJ, 1996) pp. 245Ð266.

\bibitem{kuriyan1991} 
John Kuriyan, Klara Osapay, Stephen K.~Burley, Axel T.~Br{\"u}nger, Wayne A. Hendrickson, and Martin Karplus, ``Exploration of disorder in protein structures by x-ray restrained molec- ular dynamics,''
{\it Proteins: Structure, Function, and Bioinformatics} {\bf10}, 340Ð358 (1991).

\bibitem{burling1994} 
F.~Temple Burling and Axel T.~Br{\"u}nger, 
``Thermal motion and conformational disorder in protein crystal structures: Comparison of multi-conformer and time-averaging models,''
{\it Israel Journal of Chemistry} {\bf34}, 165Ð175 (1994).

\bibitem{levin2007} 
Elena J Levin, Dmitry A Kondrashov, Gary E Wesenberg, and George N Phillips, 
``Ensemble refinement of protein crystal structures: validation and application,''
{\it Structure} {\bf15} 9, 1040Ð52 (2007).

\bibitem{woldeyes2014} 
Rahel A Woldeyes, David A Sivak, and James S Fraser, 
``E pluribus unum, no more: from one crystal, many conformations,''
{\it Current opinion in structural biology} {\bf28}, 56Ð62 (2014).

\bibitem{keedy2015} 
Daniel A Keedy, James S Fraser, and Henry van den Bedem, 
``Exposing hidden alternative backbone conformations in x-ray crystallography using qfit,''
{\it PLoS computational biology} {\bf11}, e1004507 (2015).

\bibitem{kendrew1997} 
Steven G Kendrew, David A Hopwood, and EN Marsh, 
``Identification of a monooxygenase from streptomyces coelicolor a3 (2) involved in biosynthesis of actinorhodin: purification and characterization of the recombinant enzyme,''
{\it Journal of bacteriology} {\bf179}, 4305Ð4310 (1997).

\end{thebibliography}
\end{document}